\documentclass[conference,a4paper]{APSIPA2026}
\usepackage{bm}
\usepackage{amssymb}
\usepackage{amsmath}
\usepackage{graphicx}
\usepackage{multirow}
\usepackage{threeparttable}
\usepackage[backend=biber,style=ieee,]{biblatex}
\usepackage{geometry}
\usepackage{fancyhdr}

\fancypagestyle{firststyle}{
  \fancyhf{}
  \fancyhead[C]{2026 Asia Pacific Signal and Information Processing Association Annual Summit and Conference (APSIPA ASC)}
}

\begin{document}

\title{Enhancing Neural Speech Coding with Semantic and Visual Cues}

\author{
\authorblockN{
Yao Guo,
Yang Ai$^*$,
Hui-Peng Du,
Xiao-Hang Jiang,
Chen-Yuan Ning,
Zhen-Hua Ling
}

\authorblockA{
National Engineering Research Center of Speech and Language Information Processing, \\University of Science and Technology of China, Hefei, P. R. China \\
E-mail: guoyao1917@mail.ustc.edu.cn, yangai@ustc.edu.cn, redmist@mail.ustc.edu.cn, \\jiang\_xiaohang@mail.ustc.edu.cn, chenyuanning@mail.ustc.edu.cn, zhling@ustc.edu.cn}
}

\maketitle
\renewcommand{\thefootnote}{\fnsymbol{footnote}}
\footnotetext[1]{Corresponding author. This work was funded by the National Nature Science Foundation of China under Grant 62301521.}
\renewcommand{\thefootnote}{\arabic{footnote}}
\thispagestyle{firststyle}
\pagestyle{empty}

\begin{abstract}
At low bitrates, neural speech codecs have limited capacity to encode all information needed for high-quality reconstruction, especially when relying solely on speech-derived representations.
To address this limitation, this paper proposes a Semantic- and Visual-enhanced Speech Codec (SVSC), which incorporates semantic and visual cues into the neural speech coding process. 
Specifically, built upon a mainstream neural speech coding architecture, SVSC introduces a semantic encoding-decoding branch and an image analysis-synthesis branch. 
It fuses deep semantic features with visual cues through a cross-attention mechanism, forming an auxiliary high-level representation enriched with contextual and articulatory information.
To handle different inference scenarios, SVSC introduces two information-injection strategies based on the availability of auxiliary semantic and visual cues. 
When such cues are available, the fusion mode directly incorporates the auxiliary representations into the speech coding branch through feature concatenation; otherwise, the distillation mode transfers auxiliary information into the speech coding branch through knowledge distillation during training, enabling speech-only inference without additional inputs.
Experimental results validate the effectiveness of incorporating semantic and visual cues, improving the ViSQOL score of reconstructed speech from 3.86 to 4.01.
\end{abstract}


\section{Introduction}
Speech coding aims to compress speech signals to lower bitrates while preserving perceptual quality, and it plays a critical role in applications such as audio transmission and storage.
In recent years, neural speech coding has attracted significant
attention. 
Neural speech codecs operate within an end-to-end trainable framework, leveraging neural networks to jointly design and train the encoder, quantizer, and decoder.  
One of the pioneering efforts in this direction is SoundStream \cite{zeghidour2021soundstream}, which directly processes the raw waveform through an encoder, followed by a residual vector quantizer (RVQ) \cite{vasuki2006review}, and reconstructs the speech with a decoder. 
As noted in the literature, neural audio codecs have largely surpassed traditional codecs in rate–distortion performance while supporting new use cases such as joint coding and analysis, generative modeling, and downstream integration with large language models (LLMs) and audio editing pipelines. 
Subsequent models such as Encodec \cite{defossezhigh} and HiFi-Codec \cite{yang2023hifi} can be viewed as extensions of this architecture, introducing improved training procedures and quantization schemes. 
For instance, HiFi-Codec proposes a group-residual vector quantization (GRVQ) technique that achieves strong reconstruction performance with only 4 codebooks, significantly reducing the burden on downstream generative models. 
To circumvent the challenges of direct waveform modeling and to enhance generation efficiency, APCodec \cite{ai2024apcodec} adopts a parallel framework for encoding and decoding speech amplitude and phase spectra. 
More recently, MDCTCodec \cite{jiang2024mdctcodec} has adopted a simpler, single-path structure that targets the modified discrete cosine transform (MDCT) spectrum. 
This design achieves superior generation efficiency while maintaining high reconstruction quality.

Despite these advances, low-bitrate neural speech coding remains challenging. At low bitrates, the quantized latent codes have limited capacity and must preserve phonetic content, speaker characteristics, prosody, and fine acoustic details within a highly compact representation. When the codec relies solely on speech-derived representations, some high-level contextual and articulatory information may not be sufficiently preserved, which limits the quality and naturalness of the reconstructed speech. To alleviate this information bottleneck, auxiliary high-level cues can be introduced into the neural speech coding process. Two types of auxiliary information are particularly relevant for this purpose.

\begin{figure*}
    \centering
    \includegraphics[width=0.7\linewidth]{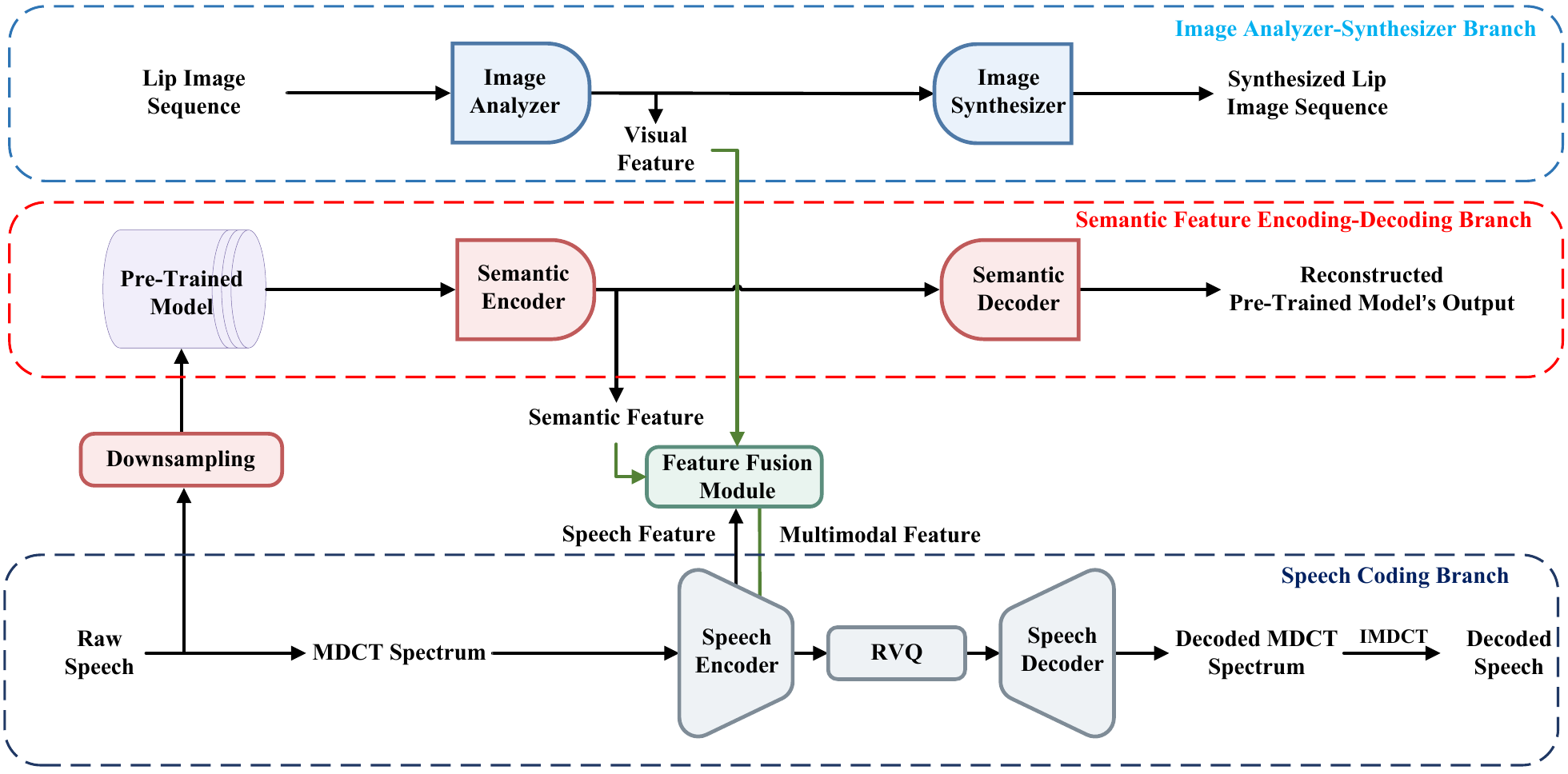}
    \caption{An overview of the proposed SVSC.}
    \label{fig:1}
\end{figure*}

The first type of auxiliary information is semantic cues derived from self-supervised pre-trained speech models, such as wav2vec 2.0 \cite{baevski2020wav2vec}, HuBERT \cite{hsu2021hubert}, and WavLM \cite{chen2022wavlm}. 
Trained on large-scale unlabeled speech data, these models can capture rich contextual and semantic information beyond local acoustic frames. Such high-level representations can provide useful guidance for neural speech coding, especially under low-bitrate conditions where the capacity of speech codes is limited. Recent neural speech codecs, such as DM-Codec \cite{ahasan2024dm} and DualCodec \cite{li2025dualcodec}, have explored knowledge distillation or feature fusion to incorporate pre-trained speech representations into codec models, leading to improved reconstruction quality.

The second type of auxiliary information comes from visual cues related to speech production, especially lip and articulatory movements. Visual information is naturally correlated with speech content and speaker characteristics, and has been widely used in speech-related tasks \cite{xu2022vsegan,xu2022improving,zheng2024incorporating}. For example, Xu \MakeLowercase{\textit{et al.}} used a multi-head cross-attention mechanism to fuse audio and lip features for audio-visual speech enhancement \cite{xu2022improving}. Zheng \MakeLowercase{\textit{et al.}} incorporated ultrasound tongue images to improve lip-based speech enhancement \cite{zheng2024incorporating}. Researchers have also developed audio-visual corpora \cite{ribeiro2021tal} and investigated audio-visual speech recognition \cite{afouras2018deep}. Inspired by these studies, our previous work \cite{guo2025vision} introduced visual information into neural speech coding and improved the quality of decoded speech.

Motivated by the complementarity between semantic and visual cues, this paper proposes a Semantic- and Visual-enhanced Speech Codec (SVSC) for low-bitrate neural speech coding. 
SVSC introduces a semantic encoding-decoding branch and an image analysis-synthesis branch to extract contextual and articulatory information, respectively, and fuses them through cross-attention to form an auxiliary high-level representation. 
To support inference scenarios with and without auxiliary cues, SVSC adopts two feature-injection strategies, i.e., direct concatenation and knowledge distillation. 
Experimental results demonstrate that SVSC consistently improves reconstruction quality over the MDCTCodec baseline, validating the effectiveness of semantic and visual enhancement.


\section{Proposed Method}

\subsection{Overview}

Fig. \ref{fig:1} illustrates the overall framework of the proposed SVSC. Building on our previous work \cite{guo2025vision}, SVSC is developed within the MDCTCodec framework \cite{jiang2024mdctcodec} to introduce semantic and visual cues for low-bitrate neural speech coding. 
Specifically, SVSC extends the speech coding branch with two auxiliary branches, i.e., a semantic encoding-decoding branch and an image analysis-synthesis branch. 
The semantic branch extracts contextual cues from a self-supervised pre-trained speech model, while the image branch extracts articulatory cues from lip image sequences. These two types of auxiliary cues are fused through a cross-attention mechanism to form a high-level auxiliary representation.

To support different inference scenarios, SVSC adopts two feature-injection modes. For audio-visual communication scenarios where auxiliary cues such as lip images are available, the fusion mode directly injects the auxiliary representation into the speech coding branch through feature concatenation. 
For conventional speech-only coding scenarios where such auxiliary cues are unavailable or inconvenient to use, the distillation mode transfers auxiliary information into the speech encoder through knowledge distillation during training, enabling speech-only inference.


\subsection{Speech Coding Branch}
Consistent with our previous work \cite{guo2025vision}, we adopt MDCTCodec \cite{jiang2024mdctcodec} as the backbone of the speech coding branch, which is recognized as an efficient and lightweight neural speech coding framework.
The speech encoder takes the MDCT spectrum of the raw speech as input, combines it with bimodal information from the feature fusion module, and outputs the encoded multimodal features. 
These features are then discretized by an RVQ. 
The speech decoder finally decodes the MDCT spectrum from the quantized results and reconstructs the waveform via inverse MDCT (IMDCT).

Specifically, the speech encoder is composed of a pre-processing module, eight cascaded modified ConvNeXt v2 (MCNX v2) blocks, and a post-processing module. At the input end of the speech encoder, feature preprocessing is performed using a simple 1D convolutional layer followed by layer normalization. At the output end, the post-processing module includes layer normalization, a linear layer, a downsampling 1D convolutional layer, and a plain 1D convolutional layer. As demonstrated in prior work \cite{ai2024apcodec,jiang2024mdctcodec}, the MCNX v2 blocks are highly suitable for speech coding tasks. Each MCNX v2 block is built around a 1D depth-wise convolutional layer, which is followed by layer normalization, a linear layer combined with a global response normalization \cite{woo2023convnext} layer, and a Gaussian error linear unit activation function \cite{hendrycks2016gaussian}, with residual connections.
The speech decoder is largely mirror-symmetric to the speech encoder, except that the downsampling 1D convolutional layer is replaced by a 1D upsampling transposed convolutional layer.

\begin{figure*}
    \centering
    \includegraphics[width=0.9\linewidth]{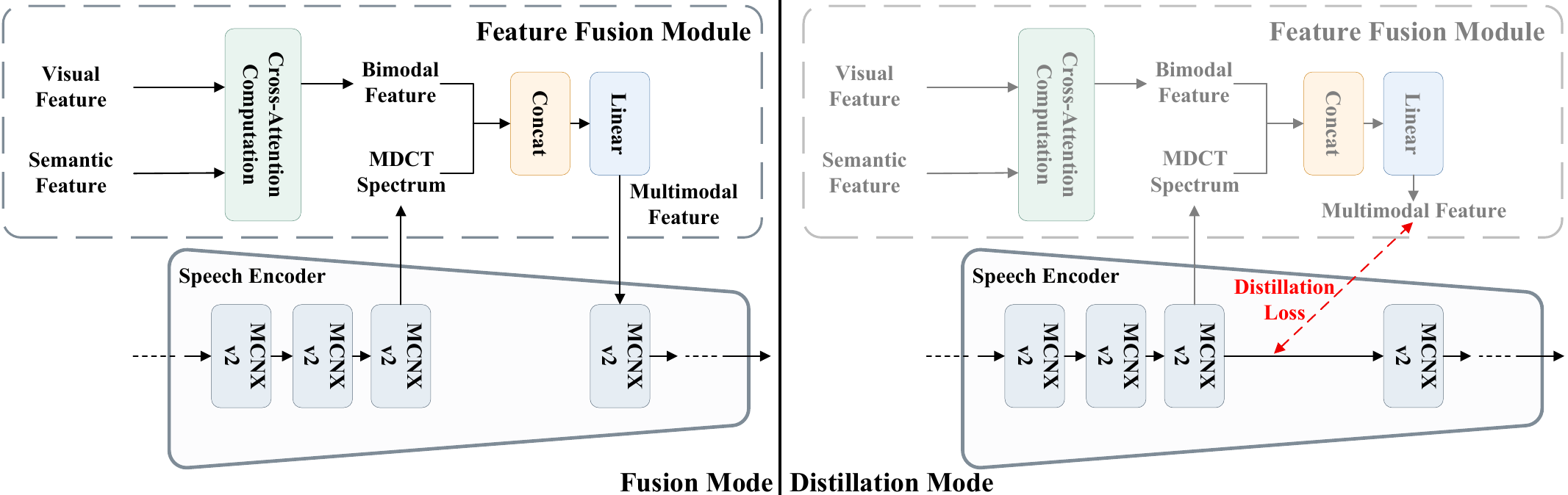}
    \caption{Overview of the two feature-injection modes in SVSC. Gray modules are used only during training and are removed during inference.}
    \label{fig:2}
\end{figure*}

\subsection{Auxiliary-Modality Feature Processing Branches}

\subsubsection{Semantic Feature Encoding-Decoding Branch}

Since the MDCT spectrum is extracted from 48-kHz speech while the self-supervised pre-trained speech model operates on 16-kHz speech, the raw speech is first downsampled to 16 kHz before being fed into the pre-trained model. 
This produces a pre-trained model's output sequence $\bm{P} \in \mathbb{R}^{D_p \times N'}$, where $D_p$ and $N'$ denote the feature dimension and the number of frames, respectively.

As shown in Fig.~\ref{fig:1}, the semantic feature encoding-decoding branch consists of a semantic encoder $\phi_{SE}$ and a semantic decoder $\phi_{SD}$. The semantic encoder transforms $\bm{P}$ into a compact semantic feature $\bm{F}'_S \in \mathbb{R}^{D_s \times N'}$, where $D_s$ denotes the dimension of the semantic feature. 
The semantic decoder then reconstructs $\bm{F}'_S$ back to the pre-trained model's output space, producing the reconstructed representation $\hat{\bm{P}} \in \mathbb{R}^{D_p \times N'}$ for semantic reconstruction loss computation. 
The above process can be summarized by the following equations:
\begin{equation}
\bm{F}_S' = \phi_{SE}(\bm{P}),\
 {\hat{\bm{P}}} = \phi_{SD}(\bm{F}_S').
\end{equation}
Specifically, the semantic encoder consists of an initial 1D convolutional layer followed by two encoder blocks. 
Each encoder block contains two residual units, and each residual unit is composed of two 1D convolutional layers with a residual connection. 
The semantic decoder is designed as a mirror-symmetric counterpart of the semantic encoder. 
To align the semantic feature with the speech coding branch, $\bm{F}'_S$ is temporally expanded by frame replication to match the number of MDCT spectrum frames. 
The final aligned semantic feature is denoted as $\bm{F}_S \in \mathbb{R}^{D_s \times N}$, where $N$ denotes the temporal length of the MDCT spectrum.
A semantic reconstruction loss $\mathcal{ L_{S}}$ is defined as the mean square error (MSE) between the reconstructed $\hat{\bm{P}}$ and the original $\bm{P}$, i.e.,

\begin{equation}
\mathcal L_{S} = \frac{1}{{D_p N'}}\mathbb E_{(\bm{P},\hat{\bm{P}})}\left\Vert \bm{P}-\hat{\bm{P}}\right\Vert_F^2.
\end{equation}
Here, $\Vert\cdot\Vert_F$ denotes the Frobenius norm.

\subsubsection{Image Analysis-Synthesis Branch}
The image analysis-synthesis branch consists of an image analyzer $\phi_{IA}$ and an image synthesizer $\phi_{IS}$. 
The image analyzer processes the input lip image sequence $\bm{I} \in {\mathbb{R}^{{H}\times{W} \times N}}$ and extracts the visual feature $\bm{F}_V \in {\mathbb{R}^{{D_v} \times N}}$, where $H$ and $W$ denote the height and width of each image, respectively, and $D_v$ represents the dimension of the visual feature. 
The image sequence $\bm{I}$ is time-aligned with the MDCT spectrum in the speech coding branch, with the temporal length being $N$. 
The image synthesizer then reconstructs the visual feature $\bm{F}_V$ into the image $\hat{\bm{I}} \in {\mathbb{R}^{{H}\times{W} \times N}}$. 
This process can be represented by the following equations:
\begin{equation}
\bm{F}_V = \phi_{IA}(\bm{I}),\
\hat {\bm{I}} = \phi_{IS}(\bm{F}_V).
\end{equation}

Specifically, the image analyzer consists of five cascaded analysis blocks and a post-processing block. The first analysis block performs spatial downsampling using a strided 3D convolutional layer, followed by batch normalization and a rectified linear unit (ReLU) activation. 
In the remaining four analysis blocks, spatial downsampling is performed by an additional pooling layer, while the 3D convolutional layer is used for feature extraction. The post-processing block consists of a linear layer and four 1D convolutional layers, which reshape the extracted visual features and reduce their dimensionality. The image synthesizer is designed as a mirror-symmetric counterpart of the image analyzer. It replaces the 3D convolutional layers with 3D transposed convolutional layers and removes the pooling layers, using the transposed convolutions to restore the spatial resolution of the lip image sequence.

During the training phase, we define an image reconstruction loss between $\bm{I}$ and $\hat{\bm{I}}$ as their MSE, i.e., 
\begin{equation}
\mathcal L_{I} = \frac{1}{{HWN}}\mathbb E_{(\bm{I},\hat{\bm{I}})}\left\Vert \bm{I}-\hat{\bm{I}}\right\Vert_F^2.
\end{equation}

\subsection{Multimodal Feature Fusion}
The feature fusion module is designed to integrate auxiliary semantic and visual cues and connect them with the speech coding branch. 
It first fuses the semantic and visual features through a cross-attention mechanism to obtain a bimodal high-level representation. 
The resulting representation is then injected into the speech encoder according to the selected feature-injection mode, i.e., direct concatenation in the fusion mode or knowledge distillation in the distillation mode.

Specifically, as shown in Fig.~\ref{fig:2}, the semantic feature $\bm{F}_S$ and the visual feature $\bm{F}_V$ are first integrated through a cross-attention mechanism before being injected into the speech coding branch. 
In the cross-attention computation, $\bm{F}_S$ is used as the key and value, while $\bm{F}_V$ serves as the query. 
The resulting output is denoted as the bimodal high-level feature $\bm{F}_b \in \mathbb{R}^{D_b \times N}$, where $D_b$ is the dimension of the bimodal feature. 
This bimodal feature is then combined with the intermediate spectral feature of the speech encoder according to the selected feature-injection mode.


When auxiliary semantic and visual cues are available during inference, SVSC adopts the fusion mode to explicitly inject the auxiliary representation into the speech coding branch. 
Following our previous work \cite{guo2025vision}, we take the output MDCT spectral feature of the third MCNX v2 block in the speech encoder as the injection point. Let this feature be $\bm{X}_3 \in \mathbb{R}^{D_M \times N}$, where $D_M$ denotes the feature dimension. The feature fusion module concatenates $\bm{X}_3$ with the bimodal feature $\bm{F}_b$ along the channel dimension, and then applies a linear layer to project the concatenated feature back to the same dimension as $\bm{X}_3$, i.e.,
\begin{equation}
    \hat{\bm{X}}_3 = {\rm Concat}(\bm{X}_3, \bm{F}_b), \quad
    \tilde{\bm{X}}_3 = {\rm Linear}(\hat{\bm{X}}_3),
\end{equation}
where ${\rm Concat}$ denotes channel-wise concatenation, $\hat{\bm{X}}_3 \in \mathbb{R}^{(D_M+D_b)\times N}$ is the concatenated feature, and $\tilde{\bm{X}}_3 \in \mathbb{R}^{D_M\times N}$ is the resulting multimodal feature. 
In the fusion mode, $\tilde{\bm{X}}_3$ is directly fed into the subsequent MCNX v2 block, allowing the speech encoder to explicitly exploit the auxiliary semantic and visual information.

When auxiliary semantic and visual cues are unavailable or inconvenient to use during inference, SVSC adopts the distillation mode to implicitly transfer auxiliary information into the speech encoder. 
During training, the same feature fusion operation is used to construct the multimodal feature $\tilde{\bm{X}}_3$, which serves as the teacher-side auxiliary representation. The speech feature $\bm{X}_3$ is then encouraged to approach $\tilde{\bm{X}}_3$ through a distillation loss, defined as
\begin{equation}
\mathcal L_D =\log\left(1 + {e^{ - \frac{{{tr(\bm{X}_3^\top}  {{\tilde{ \bm{X}}}_3)}}}{{\max (||{\bm{X}_3}{||_F},\varepsilon ) \cdot \max (||{{\tilde{\bm{X}}}_3}{||_F},\varepsilon )}}}}\right),
\end{equation}
where $\operatorname{tr}(\cdot)$ denotes the matrix trace, and $\varepsilon=1\times10^{-6}$ is a lower bound for numerical stability. 
This loss encourages the speech encoder to absorb high-level auxiliary information from the multimodal feature during training. Therefore, during inference, the auxiliary branches and feature fusion module can be removed, and the codec can perform speech-only coding without additional inputs or computational overhead.



\subsection{Training Criteria}

The SVSC is trained following the generative adversarial training strategy of MDCTCodec~\cite{jiang2024mdctcodec}. 
The speech coding branch adopts the original MDCTCodec loss, denoted as $\mathcal{L}_{\mathrm{MDCTCodec}}$. 
In addition, the semantic and visual auxiliary branches are supervised by the semantic reconstruction loss $\mathcal{L}_S$ and the image reconstruction loss $\mathcal{L}_I$, respectively.

For the fusion mode, the overall training objective is defined as
\begin{equation}
\mathcal{L}_{\mathrm{SVSC}} =
\mathcal{L}_{\mathrm{MDCTCodec}}
+ \lambda_I \mathcal{L}_I
+ \lambda_S \mathcal{L}_S .
\end{equation}
For the distillation mode, the distillation loss $\mathcal{L}_D$ is further introduced to transfer auxiliary information into the speech encoder. The training objective is therefore given by
\begin{equation}
\mathcal{L}_{\mathrm{SVSC}} =
\mathcal{L}_{\mathrm{MDCTCodec}}
+ \lambda_I \mathcal{L}_I
+ \lambda_S \mathcal{L}_S
+ \lambda_D \mathcal{L}_D ,
\end{equation}
where $\lambda_I$, $\lambda_S$, and $\lambda_D$ are hyperparameters controlling the weights of different loss terms.

\section{Experiments and Results}

\subsection{Datasets}
In the experiment, we utilized the Tongue and Lips (TaL) corpus \cite{ribeiro2021tal}, a multi-speaker dataset containing ultrasound tongue imaging, optical lip videos and speech for each utterance. 
We focused on the TaL80 subset, which included recordings from 81 native English speakers without any voice talent. 
Only the speech and lip video data were used in our experiments. 
The lip video was recorded at a frame rate of 60 fps, and the speech was recorded at a sampling rate of 48 kHz at a depth of 16 bit. 
The training and validation sets contained 11,478 and 810 utterances, respectively, while the remaining 1,940 utterances formed the test set for inference.
The content of these three sets was mutually exclusive.




\subsection{Experimental Settings}

In SVSC, the configuration of the speech coding branch follows MDCTCodec with a bitrate of 6 kbps at a sampling rate of 48 kHz~\cite{jiang2024mdctcodec}. 
The frame shift for extracting the MDCT spectrum is set to 40, resulting in a frame rate of 1.2 kHz for the speech feature $\bm{X}_3$. 
The dimension of $\bm{X}_3$ is set to $D_M=256$. 
We use WavLM~\cite{chen2022wavlm} as the self-supervised pre-trained speech model for semantic feature extraction. 
The 48-kHz raw speech waveform is first downsampled to 16 kHz and then fed into WavLM, and the output of the last Transformer layer is used as the semantic representation. 
The frame rate of the WavLM output is 50 Hz, which is then upsampled to 150 Hz after being processed by the semantic encoder to achieve temporal alignment with the visual features. 
For the visual branch, the lip videos in the TaL80 set are converted into lip image sequences with a spatial resolution of $H \times W = 64 \times 64$. 
Since the original lip video frame rate is 60 Hz, the lip image sequence is upsampled to 150 Hz using FFmpeg~\cite{tomar2006converting}. 
Each frame of the bimodal features, which integrate semantic and visual information, is further replicated eight times to reach the frame rate of 1.2 kHz, enabling fusion with the intermediate MDCT spectral feature $\bm{X}_3$.

For the semantic feature encoding-decoding branch, the WavLM output dimension is set to $D_p=768$. 
The initial 1D convolutional layer uses a kernel size of 3, a stride of 1, and no bias, mapping the feature dimension from 768 to 256. 
The semantic encoder contains two encoder blocks, and each encoder block includes two residual units. 
Each residual unit is implemented with two 1D convolutional layers and a residual connection, with the feature dimension kept at 256. 
A linear layer is then used to project the feature dimension from 256 to 64, resulting in the semantic feature dimension $D_s=64$. 
For the image analysis-synthesis branch, all parameter settings follow our previous work~\cite{guo2025vision}. 
The image analyzer outputs a 64-dimensional visual feature, i.e., $D_v=64$. 
After cross-attention, the resulting bimodal feature dimension is set to $D_b=64$. 

We trained SVSC on a single NVIDIA RTX 4090D GPU with a batch size of 16. 
The loss weights were set to $\lambda_I = 0.5 \times 10^{-6}$, $\lambda_S = 1000$, and $\lambda_D = 1$.
We employed the AdamW optimizer for training, with exponential decay rates $\beta_1 = 0.8$ and $\beta_2 = 0.99$. 
The learning rate was initialized to $2 \times 10^{-4}$ and decayed by a factor of 0.999 after each epoch.

\begin{table}[t]
\centering
\caption{Evaluation results of SVSC and compared advanced neural speech codecs on the test set.}
\label{Table 1}
\setlength{\tabcolsep}{0.4mm}{
\begin{tabular}{c|cccccc}
\hline
            & PESQ          & CSIG          & CBAK          & COVL          & STOI          & ViSQOL        \\ \hline
SVSC (Fusion Mode)    & \textbf{3.37} & \textbf{4.89} & \textbf{3.54} & \textbf{4.19} & \textbf{0.96} & \textbf{4.01} \\
SVSC (Distillation Mode)   & 3.29          & 4.82          & 3.50          & 4.10          & 0.95          & 3.95          \\
MDCTCodec   & 3.25          & 4.77          & 3.28          & 4.06          & 0.95          & 3.86          \\
Encodec     & 2.95          & 4.41          & \textbf{3.54} & 3.70          & 0.92          & 3.20          \\
SoundStream & 2.64          & 4.20          & 3.39          & 3.46          & 0.90          & 3.07          \\
HiFi-Codec  & 3.27          & 4.74          & 3.52          & 4.06          & 0.93          & 3.55         \\ \hline
\end{tabular}}
\end{table}

\subsection{Baselines and Evaluation Metrics}

We compare SVSC with several representative neural speech codecs, including SoundStream~\cite{zeghidour2021soundstream}, Encodec~\cite{defossezhigh}, HiFi-Codec~\cite{yang2023hifi}, and MDCTCodec~\cite{jiang2024mdctcodec}. 
Among them, MDCTCodec is used as the direct backbone baseline, since SVSC is developed within the MDCTCodec framework. 

We adopt PESQ~\cite{rec2007p}, STOI~\cite{taal2010short}, ViSQOL~\cite{chinen2020visqol}, and three composite metrics~\cite{hu2006evaluation}, i.e., CSIG, CBAK, and COVL, to evaluate the quality of reconstructed speech. 
PESQ and ViSQOL are used to measure perceptual speech quality, STOI is used to evaluate speech intelligibility, and CSIG, CBAK, and COVL are used to assess signal distortion, background intrusiveness, and overall quality, respectively. 
For all these metrics, higher values indicate better reconstruction quality.


\subsection{Comparison with Advanced Codecs}
As shown in Table \ref{Table 1}, the comparison results between SVSC and advanced neural codec baselines demonstrate the effectiveness of the proposed semantic and visual enhancement. Compared with the MDCTCodec backbone, SVSC achieves consistent improvements in both fusion and distillation modes. In the fusion mode, where auxiliary cues are available during inference, SVSC obtains the best or tied-best performance on most objective metrics. Specifically, it improves PESQ from 3.25 to 3.37 and ViSQOL from 3.86 to 4.01, indicating that the explicitly injected auxiliary representation provides useful guidance for high-quality speech reconstruction.

The distillation mode of SVSC also improves over MDCTCodec on most metrics while requiring only speech input during inference. It increases the ViSQOL score from 3.86 to 3.95 and improves PESQ, CSIG, CBAK, and COVL compared with the backbone. These results suggest that knowledge distillation can transfer useful semantic and visual information into the speech encoder during training, enabling speech-only inference without additional auxiliary inputs.

Comparing the two modes, the fusion mode performs better than the distillation mode, which is expected because it directly uses auxiliary semantic and visual cues during inference. In contrast, the distillation mode relies on auxiliary information implicitly learned during training. This gap reflects the trade-off between reconstruction quality and inference practicality: the fusion mode provides better quality when auxiliary cues are available, while the distillation mode offers a deployable speech-only codec when such cues are unavailable or inconvenient to use.

Overall, SVSC achieves competitive or superior performance across different objective metrics. These results show that semantic and visual cues can help alleviate the information bottleneck of low-bitrate speech coding and improve the reconstruction quality of neural speech codecs.

\begin{table}[t]
\centering
\caption{Evaluation results of ablation studies.}
\label{Table 2}
\setlength{\tabcolsep}{0.4mm}{
\begin{tabular}{c|cccccc}
\hline
             & PESQ          & CSIG          & CBAK          & COVL          & STOI          & ViSQOL        \\ \hline
SVSC (Fusion Mode) & \textbf{3.37} & \textbf{4.89} & \textbf{3.54} & \textbf{4.19} & \textbf{0.96} & \textbf{4.01} \\
w/o Visual                                           & 3.34          & 4.86          & 3.50          & 4.16          & 0.96          & 3.98          \\
 w/o Semantic                                                  & 3.28          &{4.79}          &{3.32}          &{4.05}          &{0.94}          &{3.85}    \\ \hline      
\end{tabular}}
\end{table}

\subsection{Ablation Studies}

To investigate the contribution of different auxiliary cues, we conduct ablation studies by removing one type of auxiliary information from SVSC. For simplicity, the ablation experiments are performed only in the fusion mode, since this mode directly uses auxiliary cues during inference and can more clearly reflect the effect of each cue. Specifically, we compare the full SVSC with two variants: one removes the visual cues and uses only semantic information, denoted as w/o Visual; the other removes the semantic cues and uses only visual information, denoted as w/o Semantic.

As shown in Table II, the full SVSC achieves the best performance across all objective metrics, indicating that semantic and visual cues provide complementary information for low-bitrate speech reconstruction. Removing visual cues leads to a performance drop, with the ViSQOL score decreasing from 4.01 to 3.98. A larger degradation is observed when semantic cues are removed, where the ViSQOL score decreases to 3.85 and other metrics also decline. These results suggest that both auxiliary cues are beneficial, and their joint use through cross-attention provides more effective guidance than using either cue alone.

\section{Conclusion}

This paper proposed SVSC, a semantic- and visual-enhanced speech codec for low-bitrate neural speech coding. By incorporating semantic cues from WavLM and visual cues from lip image sequences, SVSC introduces auxiliary contextual and articulatory information into the speech coding process. To support different inference scenarios, SVSC adopts two feature-injection modes, namely direct feature concatenation in the fusion mode and knowledge distillation in the distillation mode. Experimental results on the TaL80 dataset show that SVSC improves reconstruction quality over the MDCTCodec backbone and achieves competitive performance against representative neural speech codecs. Ablation studies further confirm that semantic and visual cues are complementary, and their joint use brings better performance than using either cue alone. Future work will investigate more efficient auxiliary-information transfer and robustness under diverse acoustic and visual conditions.

\printbibliography

\end{document}